%% file: paper.tex
\documentclass[10pt,sigconf,letterpaper]{acmart}

\usepackage{balance}
\usepackage{subcaption}[tight]
\usepackage{fancyvrb}
\usepackage{xspace}
\usepackage{enumitem}
\usepackage{algpseudocode}
\usepackage{algorithm}
\algtext*{EndIf}

\newcommand{\ie}{i.e., \@}
\newcommand{\eg}{e.g., \@}

\newcommand{\etal}{et al.\xspace}

\newcommand{\punkt}[1]{\item\textbf{#1}:}
\newcommand{\afrinic}{AFRINIC\xspace}

\graphicspath{{figures/}}

\begin{document}

\title{IP Neo-colonialism: Geo-auditing RIR Address Registrations}

\author{Robert Beverly}
\affiliation{%
  \institution{CMAND}
  \country{}
}
\email{rbeverly@cmand.org}


\setcopyright{none}
\settopmatter{printacmref=false, printccs=false, printfolios=false}
\renewcommand\footnotetextcopyrightpermission[1]{}
\pagestyle{plain}
\acmConference{}{}{}
\renewcommand{\shortauthors}{}


%


\begin{abstract}
Allocation of the global IP address space is under the purview of
IANA, who distributes management responsibility among five
geographically distinct 
Regional Internet Registries (RIRs).  
Each RIR
is empowered to bridge technical (e.g., address uniqueness and
aggregatability) and policy (e.g., contact information and IP 
scarcity)
requirements unique to their region.  
While different RIRs have different policies for
out-of-region address use, little prior systematic analysis has 
studied \emph{where} addresses are used
post-allocation.


In this preliminary work, we e
IPv4 prefix registrations 
across the five RIRs (50k total prefixes) 
and utilize the Atlas distributed
active measurement infrastructure to geolocate
prefixes at RIR-region granularity.  We define a taxonomy
of registration ``geo-consistency'' by comparing
a prefixes' inferred physical location to the allocating
RIR's coverage region
as well as the registered organization's location.  We
then apply this methodology and taxonomy to audit the
geo-consistency of 10k
random IPv4 prefix allocations within each RIR (50k total
prefixes).
While we find registry information 
to largely be consistent with our geolocation
inferences, we show that some RIRs have a non-trivial fraction of
prefixes that are used both outside of the RIR's region and outside of
the registered organization's region.  A better understanding
of such discrepancies can increase transparency for the community and
inform ongoing discussions over in-region address use and policy.
\end{abstract}

\maketitle

\input{intro}
\input{background}
\input{data}
\input{method}

\input{results}
\input{concl}

\section*{Acknowledgements}
We thank RIPE for maintaining the Atlas measurement infrastructure.
John Curran, Thomas Krenc, Anita Nikolich, and Eric Rye provided
invaluable early feedback.  


\label{page:end_of_main_body}

\bibliographystyle{ACM-Reference-Format}
\bibliography{neo}

\appendix
\input{alg}

\label{page:last}



\end{document}

%% file: intro.tex
\section{Introduction}
\label{sec:intro}

To ensure the global uniqueness of public Internet Protocol (IP)
addresses, the Internet Assigned Numbers Authority (IANA) allocates
large contiguous blocks of addresses to 
Regional Internet
Registries (RIRs) who then further suballocate within their respective
geographic region.  In addition to this coordination, the
RIRs strive to make allocations efficient such that the space
is well-utilized; this is especially important in IPv4 where addresses
are a scarce resource.  Indeed, with the advent of IPv4 address
exhaustion \cite{richter2015primer}, IPv4 addresses have become a valuable commodity
\cite{prehn2020wells}.

Assignments, allocations, and reassignments are governed by RIR
policies that are designed to provide uniqueness, efficiency, and
accountability.  There are five RIRs, each of which serves a different
geographical region of the world.  
The purpose of regional delegation 
is explicitly codified by IANA: 
``RIRs are established and authorized by
respective regional communities, and recognized by the IANA to serve
and represent large geographical regions'' \cite{arin-nrpm}.

While prior work has examined inter-RIR address transfers that
are logged and public~\cite{2017-livadariu-itmw}, there has been little systematic effort
to understand the true physical location and region where addresses are
advertised once allocated.  Our goal is to: i) increase transparency
and help the community better understand where addresses are 
being used; ii) quantify the extend to which registry information
is accurate and can serve operational needs; and iii) inform
ongoing discussion and debate over out-of-region address
use and policy, e.g., ~\cite{afrinic-faq, afrinic-court, ci-seized,
sa-heist, krebs}.

Therefore, in this work we examine
the IPv4 assignments of the five RIRs and characterize the countries and 
regions of the organizations to which addresses are given.  We then
use a distributed active measurement platform
with multiple nodes in each region to perform region-granularity 
geolocation.  
We define a taxonomy of registration geo “consistency” by comparing a
prefixes’ inferred physical location to the allocating RIR’s
coverage region as well as the registered organization’s location. 
%
Our contributions include:
\begin{enumerate}[leftmargin=*]
 \item Macro-characterization of the registered IP
       addresses
       across the five RIRs, including prefix granularity,
       inter-regional registration, and pathologies.
 \item A taxonomy of prefix registration geo-consistency.
 \item An active measurement 
       campaign to determine
       whether prefixes are used in a physical location 
       consistent with the RIR's region and registered 
       organization's country.
 \item A case study of address registration
       inconsistency.
 \item Public availability of code and measurement data 
       without restriction to facilitate reproducibility.
\end{enumerate}
The remainder of the paper is organized as follows.  We provide
background on IP address allocation and RIRs along with related work
in \S\ref{sec:background}.  Next we characterize 
the prefix
registration information in the RIRs
in \S\ref{sec:data}.
We then
present our methodology for 
geo-auditing IP 
allocations in \S\ref{sec:method} and
preliminary results in \S\ref{sec:results}.
Finally, we conclude with suggestions
for future research.



%
%
%

%
%
%
%
%

%% file: background.tex
\begin{table*}[t]
\caption{NRO Comparative Policy Overview: membership and out-of-region
policies across RIRs \cite{nro}}
\label{tab:nro}
\centering
\begin{tabular}{|l p{0.8\linewidth}|}\hline
ARIN & ``\textit{To receive resources, ARIN requests organizations to
verify that it plans on using the resources within the ARIN region}''
\\\hline
RIPE & ``\textit{The network that will be using the resources must have
an active element located in the RIPE NCC service region}''
\\\hline
APNIC & ``\textit{permits account holders located within the APNIC
service region to use APNIC-delegated resources out of region}''
\\\hline
LACNIC & ``\textit{requires organizations to be legally present and have
network infrastructure in the LACNIC service region to apply for and
receive resources}''
\\\hline
AFRINIC & ``\textit{requires organizations/persons to be legally present
and the infrastructure from which the services are originating must be
located in the AFRINIC service region}''
\\\bottomrule
\end{tabular}
\end{table*}

\section{Background}
\label{sec:background}

Internet Protocol (IP) addresses are fundamental to the Internet's
operation.  The Internet numbers registry system has three primary
goals for IP addresses: 1) allocation pool management; 2) hierarchical
allocation; and 3) registration accuracy~\cite{rfc7020}.  The IP
address hierarchy is rooted in the Internet Assigned Numbers Authority
(IANA) which is managed by the Internet Corporation for Assigned Names
and Numbers (ICANN) organization.  RFC 1366 first proposed to
geographically distribute the registry functionality~\cite{rfc1366}.
Today IANA allocates large (\ie /8), unique blocks of IP address space to the
Regional Internet Registries (RIRs), of which there are currently
five: ARIN (North America),
RIPE (Europe), APNIC (Asia and Pacific), LACNIC (Latin America), or
\afrinic (Africa).  This distribution affords autonomy to the
different RIRs to consider region-specific geopolitical policies and
constraints.

Each RIR maintains registration information, including for example the
assigned organization, mailing address, and points of contact, for
numbered resources (Figure~\ref{fig:record} provides an example).  
Accurate registration information is important to
the operation and management of the highly distributed and autonomous
Internet.  Registry
information is exposed via a public directory service known as ``whois
~\cite{rfc3912}''.

Portions of address allocations are frequently assigned to
end-users, \eg an ISP's customer.  Registries such as ARIN require
the reassignment of prefixes of /29 or more addresses to be 
registered via a directory services system such as Shared Whois
(SWIP) within seven days~\cite{rfc1491,ripeswip}.  Thus, the registered prefixes
are both frequently small and up-to-date.  


\begin{figure}[!bpt]
\centering
{\scriptsize
\begin{Verbatim}[frame=single]
NetHandle:      NET-104-148-63-0-1
OrgID:          C05266659
Parent:         NET-104-148-0-0-1
NetName:        WEB-OMEGA-DO-BRASIL
NetRange:       104.148.63.0 - 104.148.63.255

OrgID:          C05266659
OrgName:        Web Omega do Brasil
Street:         Rua do Xareu, qd 13, lote 20
City:           Goiania
State/Prov:     GO
Country:        BR
\end{Verbatim}
}
\vspace{-3mm}
\caption{Example whois prefix registration record from ARIN.  The
corresponding OrgID record includes the
organization's country code (BR).  Brazil
is in the region of a different RIR, LACNIC. 
}
\vspace{-3mm}
\label{fig:record}
\end{figure}

\subsection{Motivation}

As noted in the ARIN Number Resource Policy Manual: ``The primary role of RIRs is
to manage and distribute public Internet address space within their
respective regions'' \cite{arin-nrpm}.  However, RIRs must balance
efficient and equitable use of IP addresses with the true need for
addresses, as well as real-world operational constraints.  Further, 
the different RIRs have \emph{different} policies with respect to 
out-of-region address use; Table~\ref{tab:nro} summarizes pertinent
policies from NRO's Comparative Policy Overview~\cite{nro}.  

ARIN, for instance, requires requests for address allocations to be
motivated by need within their service region, however subsequent
use outside of the region is permitted.  APNIC
explicitly allows out-of-region use without restriction.  However,
AFRINIC has a more restrictive policy that requires use in-region.
Indeed, significant debate and even legal action has been the result of
AFRINIC actions to enforce in-region address 
use~\cite{afrinic-court,afrinic-faq,ci-seized,sa-heist,krebs}.

In recognition of these complexities, we focus on 
increasing transparency by uncovering out-of-region address
use that can only be uncovered via measurement.  We further
take a conservative view of out-of-region use by 
considering both the allocating RIR and the registered
organization's location, and identifying instances where
registrations are geo-inconsistent.  Our goal is thus to
help the community 
better understand where addresses are being used post-allocation,
whether registration information is accurate and can serve operational
needs, and inform ongoing discussion over in-region address use
and policy.

\subsection{Related Work}

While the present work focuses on IPv4 addresses, the whois protocol
is also used to register contact information for domain names, \eg by
domain name registrars for various portions of the DNS namespace.
Significant prior work has examined domain names and their associated
registration information.  Liu \etal show that the domain name
registry information does not follow a consistent schema, and devise
robust parsing techniques using statistical
models~\cite{liu2015learning}.  We similarly find idiosyncrasies in the
conventions used by different RIRs, but with only five RIRs, are able
to manage these complications using fairly straightforward rules.

Prior work shows that domain names are frequently used for abusive or
malicious purposes and registration behavior can be indicative of such
misuse~\cite{leontiadis2014empirically}.  Lauinger \etal analyze the
re-registration of domain names after their expiration and show how
attackers can leverage residual trust by capturing these expired
domains~\cite{lauinger2016whois}.  More recently, Lu \etal dive into
whois in the era of GDPR to better understand domain registration
privacy~\cite{lu2021whois}.  Although these prior works all utilize
whois registration data, they all focus on domain names rather than
IP address block allocations.

The scarcity of IP addresses has created markets where addresses
can be transferred and sold as a commodity.  Livadariu analyzed
transfers published by the RIRs to characterize the size of prefixes
and their eventual use as evidenced in the global BGP
table~\cite{2013-livadariu-fltm, 2017-livadariu-itmw}.  In contrast,
we focus on understanding the true location where addresses are
used, and whether these locations are within or outside of the
regions for which the corresponding registry is responsible.  While
unofficial ``under-the-table'' transfers where the correct location
and registry information is not properly updated may explain some
of the location inconsistencies we discover, we leave causal 
analysis to future work.

%% file: data.tex
\section{Prefix Registration Analysis}
\label{sec:data}

We use raw dumps of the IPv4 registration databases as of 
April, 2023 from
all five RIRs using their ``bulk whois'' facility; this removes any
dependence on the public whois interface and ensures that we
obtain the complete data~\cite{bulkwhois}.  These data are flat text files
containing records that consist of key-value pairs, \eg ``NetRange''
and ``OrgID'' in Figure~\ref{fig:record}.  Each RIR's database has different
schemas and idiosyncrasies, these include for example the address prefix 
representation, different key names, transferred prefixes, and 
prefixes from other RIRs.  We parse each prefix allocation, skip
prefixes the RIR does not manage (but still has listed in their
database for completeness \eg with a ``not-managed-by'' note), and
map organization entries to prefixes.  This organizational mapping
allows us to tie a prefix registration's ``OrgID'' to the country
of the registered organization corresponding to
that identifier.   

We ignore pathologies in the RIR data, the most
common of which are circular references.  For example, an RIR may
list a prefix as transferred to a different RIR, while that RIR
lists the same prefix as belonging to the original RIR.  While these
types of errors could be due to time differences in the data dumps,
the ``Updated'' timestamps on the records suggest they are 
simply errors that we cannot resolve.

\begin{figure}[t] 
 \centering
 \resizebox{0.9\columnwidth}{!}{\includegraphics{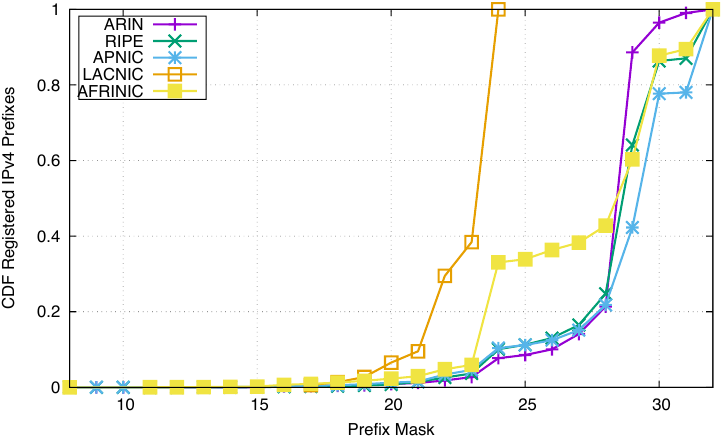}}
\vspace{-3mm}
 \caption{Distribution of registered IPv4 prefixes per-RIR as a function of prefix
length.}
 \label{fig:plen}
\end{figure}

In total, we find over 8M prefix registrations across the five RIRs,
representing a significantly more granular allocation than exists in
the global BGP table.  Note that our goal is not to compare RIR and
BGP allocations, but rather to emphasize that prefixes within the
RIRs are significantly more fine-grained.  
Figure~\ref{fig:plen} depicts the cumulative
fraction of prefixes as a function of the prefix mask across the RIRs.
We see that nearly 90\% of the registered prefixes within ARIN, RIPE
and APNIC have a prefix length of /25 or greater, suggesting active
use of SWIP.  In contrast, a larger proportion of registered prefixes 
within \afrinic and 
LACNIC are larger (smaller mask).

\begin{figure*}[t] 
 \centering
\begin{subfigure}{0.5\textwidth}
  \centering
  \includegraphics[width=.8\linewidth]{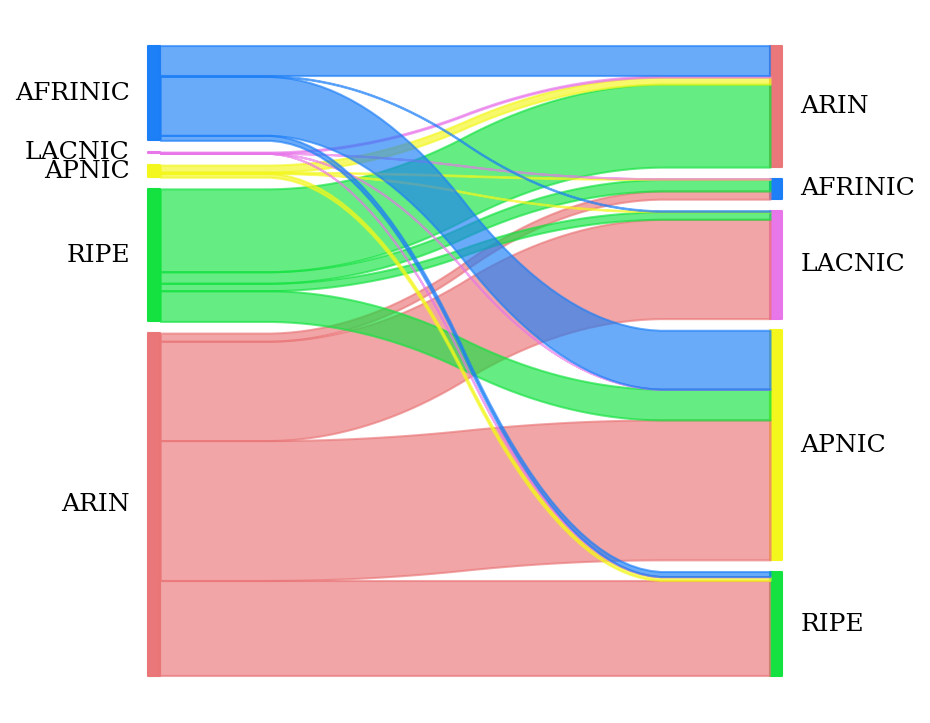}
  \vspace{-2mm}
  \caption{Inter-region registrations (prefixes)}
  \label{fig:irp}
\end{subfigure}%
\begin{subfigure}{.5\textwidth}
  \centering
  \includegraphics[width=.8\linewidth]{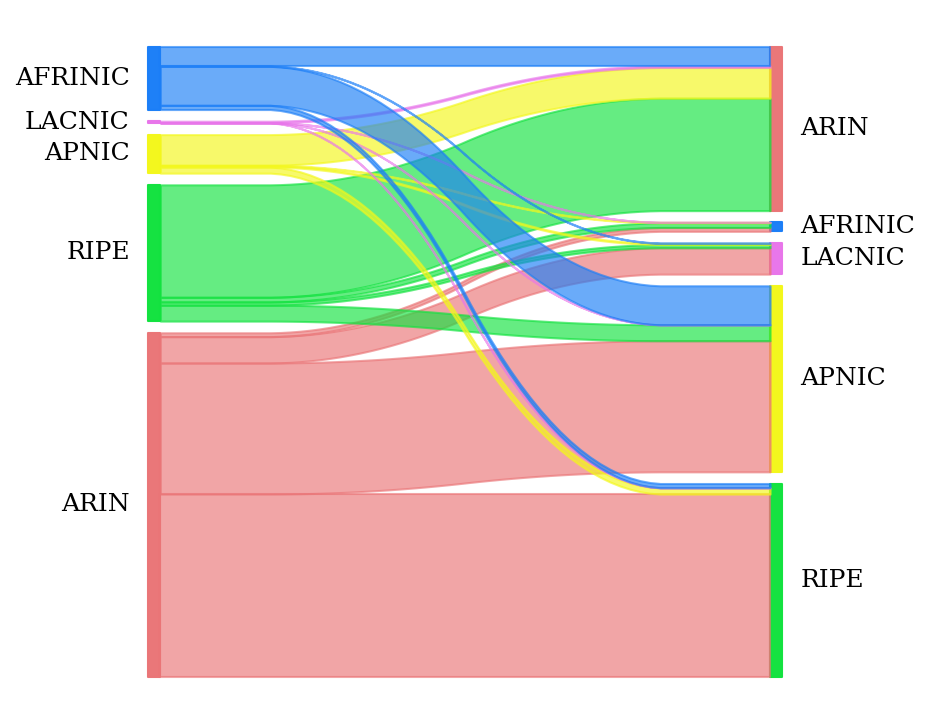}
  \vspace{-2mm}
  \caption{Inter-region registrations (addresses)}
  \label{fig:irb}
\end{subfigure}
  \vspace{-3mm}
 \caption{Inter-region registrations: proportions of prefixes (left)
and addresses (right) allocated by an RIR to an organization with a
physical address in a region for which a different RIR is
responsible.}
 \label{fig:ir}
\end{figure*}

\subsection{Inter-region Registration}
\label{sec:interregion}

We note that it is not uncommon for a registry to allocate a prefix to
an organization that is outside of the RIR's region.  For example, a
prefix $P$ belongs to an address block allocated to ARIN, but the
registered organization within the ARIN database has a physical
address in Great Britain.  In this case, the organization's country
belongs to RIPE -- a different RIR.  Figure~\ref{fig:record} provides
one such example whois record.

Indeed, as part of an effort to ensure flexibility and efficient use
of addresses, out of region registrations are permitted by RIR
policies under certain conditions.  For instance, the ARIN Number
Resource Policy Manual (NRPM) \cite{arin-nrpm} explicitly states that ``ARIN
registered resources may be used outside the ARIN service region...
provided that the applicant has a real and substantial connection with
the ARIN region.''  Other RIRs have similar policies, and
quantitatively define the requirements for out-of-region use (\eg 
at least a /22 used in region and an in-region peering session).

\begin{table}[t]
\caption{RIR prefix registration macro-statistics}
\label{tab:overview}
\vspace{-2mm}
{\small
\resizebox{\columnwidth}{!}{
\begin{tabular}{lrrrr}\hline
RIR & Prefixes & Out-region   & Addresses & Out-Region \\
    & (k)      & Prefixes (k) & (/24s)     & Addresses (/24s) \\
\midrule

ARIN & 3,109.8 & 77.3 (2.5\%) & 5,491,682 & 128,546 (2.3\%)\\
RIPE & 3,556.7 & 29.8 (0.8\%) & 2,925,866 & 50,579 (1.7\%)\\
APNIC & 1,150.8 & 2.7 (0.2\%) & 9,136,159 & 14,327 (0.2\%)\\
LACNIC & 66.5 & 0.3 (0.5\%) & 251,088 & 651 (0.3\%)\\
AFRINIC & 148.5 & 21.1 (14.2\%) & 486,456 & 23,601 (4.9\%)\\
\midrule
Total: & 8,032.3 & 131.3 & 18,291,251 & 217,705\\
\bottomrule
\end{tabular}}}
\vspace{-6mm}
\end{table}

\begin{figure*}[t!] 
 \centering
 \resizebox{1.3\columnwidth}{!}{\includegraphics{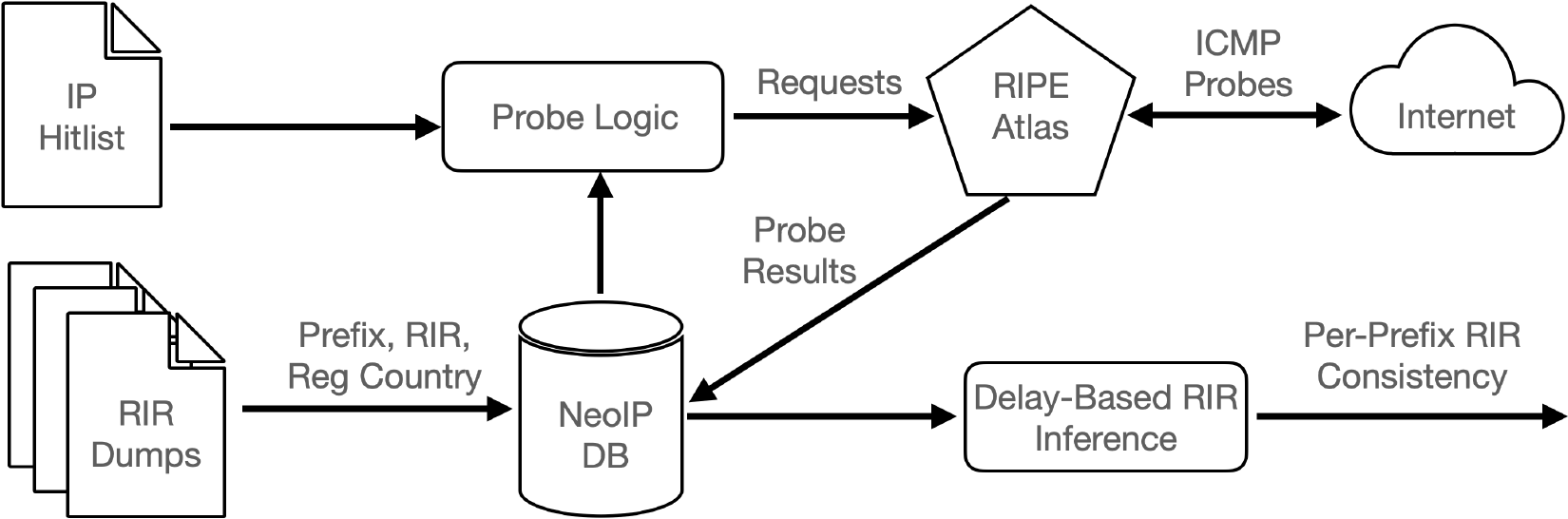}}
 \caption{RIR Geo-audit methodology.  $\sim$8M prefixes are ingested
into a database and, with a global IPv4 hitlist, used to drive the probing 
logic (what target and what vantage points).  RIPE Atlas performs
active probes used to make delay-based RIR inferences.}
 \label{fig:method}
\end{figure*}

\begin{figure*}[t!] 
 \centering
 \resizebox{1.4\columnwidth}{!}{\includegraphics{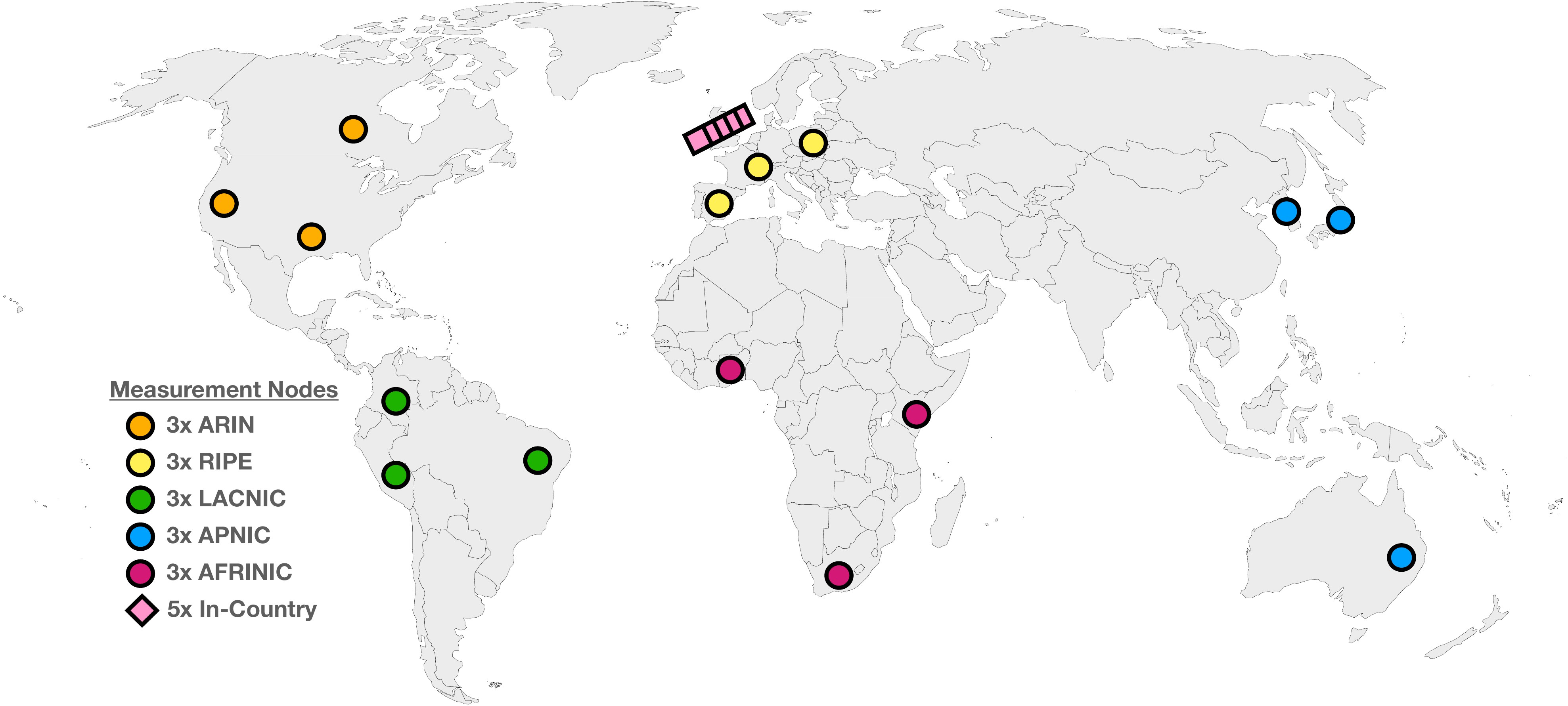}}
 \caption{Regional prefix geolocation using RIPE Atlas nodes
 for active  measurement.  Three different nodes
 within each RIR's region (total of 15 nodes) help infer whether a
 target prefix is within the assigned region.  An additional five
 nodes in the country of the prefix's registered
 organization seek to determine whether the registration 
 is consistent with the RIR responsible for that country.  If the
 inferred geolocation matches \emph{either} the responsible RIR or the 
 RIR of the organization, we call the registration ``consistent.''}
 \label{fig:atlas}
\end{figure*}

Table~\ref{tab:overview} provides a macro-level analysis of 
the number of prefixes (in thousands) in each RIR, as well as the fraction
of prefixes and addresses (in /24 equivalents) registered to
out-of-region organizations.  
Note that 
the sum of all addresses across the RIRs is larger than the total
IPv4 address space -- this is due to overlap between prefixes within
different RIRs.  For example, \afrinic is responsible for 154.0.0.0/8,
however ARIN has a registered allocation of 154.1.0.0/16 (for the
company Goldman Sachs).


ARIN has an appreciable fraction (2.5\%) of prefixes with
out-of-region registered organizations, and dominates in 
total volume of both out-of-region registered prefixes and addresses.
Given the long history of ARIN and efforts to redistribute
addresses, this finding is not wholly unsurprising.  

However, \afrinic is the most recently formed registry (2004). 
More than 14\% of \afrinic prefixes are allocated to
out-of-region organizations, suggesting significant 
apparent involvement of organizations outside of Africa and
possibly use of these resources 
outside of Africa (a hypothesis we
test in the next section).  Similarly, \afrinic leads in out-of-region
registrations when accounting by total IP addresses (4.9\%).

Figure~\ref{fig:ir} provides a Sankey diagram of the inter-region
registration activity.  Most of the out-of-region \afrinic
registrations are from organizations in Asia and America.
Interestingly, approximately the same number of ARIN prefixes are 
allocated to European organizations as RIPE prefixes are allocated 
to American organizations.


%% file: method.tex
\section{Prefix Registration Geo-Audit}
\label{sec:method}

Having examined the information available for prefixes in the RIRs, we
next turn to validating the corresponding location information.
Figure~\ref{fig:method} provides an overview of the geo-audit
methodology.  There are three primary components to our methodology: 
1) ingesting RIR prefixes into a database; 2) active
ICMP probing, including selecting the origin vantage point and 
destination targets
within prefixes; and 3) delay-based RIR inferences. Important
supporting components of this high-level methodology include:

\ifx
\begin{table}[t] 
\caption{Our geo-audit results in one of five possible outcomes; the
first two outcomes reflect \emph{consistency} with the registered prefix, 
while the remaining three results indicate a form of
\emph{inconsistency}.  
We provide an example of each result scenario for clarity.} 
\label{tab:terminology}
{\small 
\resizebox{\columnwidth}{!}{
\begin{tabular}{lrrr} 
                    &      \multicolumn{3}{c}{Example} \\ 
Result              &      $RIR_{Reg}$  &  $RIR_{Org}$  & $RIR_{Geo}$ \\ 
\midrule 
($FC$) Fully Geo-consistent      &      ARIN &  ARIN    & ARIN \\ 
($OC$) Organization Geo-consistent    &      RIPE &  ARIN    & ARIN \\ 
\midrule 
($OI$) Organization Geo-inconsistent  &      ARIN &  RIPE    & ARIN \\ 
($RI$) Registry Geo-inconsistent &      ARIN &  ARIN    & RIPE \\ 
($FI$) Fully Geo-inconsistent    &      ARIN &  RIPE    & APNIC  \\ 
\bottomrule
\end{tabular}}} 
\end{table}
\fi


\begin{itemize}[leftmargin=*]
\punkt{Mapping countries to RIRs} As a first step, we require a
mapping between countries, and their corresponding ISO country codes,
to the responsible RIR.  Using publicly available information for each
registry, we manually map 244 different countries (including dependent
territories with their own ISO code) to the five RIRs.
ARIN, RIPE, APNIC, LACNIC, and \afrinic are each currently responsible
for 29, 73, 54, 31, and 57 countries respectively.

\punkt{Prefix handling} Throughout this work, we respect IP prefixes
and their properties.  We construct a radix trie on the registered
prefixes such that we can perform longest prefix matching of an
address to its corresponding registered prefix.  Further, when 
computing statistics, we take into account subnetting and
prefix aggregation such that presented address statistics do not
double or over count.

\begin{table*}[t!] 
\caption{Our geo-audit results in one of five possible outcomes; the
first two outcomes reflect \emph{consistency} with the registered prefix, 
while the remaining three results indicate a form of
\emph{inconsistency}.  
We provide an example of each result scenario for clarity.} 
\label{tab:terminology}
\vspace{-2mm}
{\small 
\resizebox{1.9\columnwidth}{!}{
\begin{tabular}{l l|rrr} 
                    &      & \multicolumn{3}{c}{Example} \\ 
Result              & Description & $RIR_{Reg}$  &  $RIR_{Org}$  & $RIR_{Geo}$ \\ 
\midrule 
($FC$) Fully Geo-consistent          &      
Geolocates in RIR and organization's region
&ARIN &  ARIN    & ARIN \\ 
($OC$) Organization Geo-consistent   &      
Geolocates outside of RIR region, but within organization's region
&RIPE &  ARIN    & ARIN \\ 
\midrule 
($OI$) Organization Geo-inconsistent &      
Geolocates in RIR's region, but organization is out-of-region
&ARIN &  RIPE    & ARIN \\ 
($RI$) Registry Geo-inconsistent     &      
RIR and organization's region consistent, but geolocates out-of-region
&ARIN &  ARIN    & RIPE \\ 
($FI$) Fully Geo-inconsistent        &      
RIR, organization, and geolocation all in different regions
&ARIN &  RIPE    & APNIC  \\ 
\bottomrule
\end{tabular}}}
\end{table*}

\punkt{Target addresses} To perform delay-based active IP geolocation,
our method requires a responsive address within a prefix.  We utilize
the January 2023 data of a popular Internet-wide IPv4 hitlist for this
purpose~\cite{isihitlist}.  The hitlist includes a score for each 
address; to obtain address likely to respond, we filter the hitlist to 
only use address with a score above 99.  We then 
longest-prefix map IPv4 hitlist addresses
to their respective RIR prefix
to generate candidate targets within fine-grained RIR prefixes 
for active probing.

\punkt{Active Probing} To perform delay-based IP geolocation, we 
require vantage points within the country where the prefix is
registered. For this purpose, we utilize the rich coverage afforded
by the RIPE Atlas project~\cite{bajpai2015survey}.  As of this writing, Atlas
has over 43k ``probes'' (vantage points) with coverage in 
over 87\% of all
world countries~\cite{atlas}. 
Each Atlas probe includes meta-data with its physical country
location.  We discuss limitations of latency-based IP geolocation and
how we minimize their effects in~\S\ref{sec:method:delay}.

\punkt{Anycast} A potential source of error in our geo-audit is 
anycast~\cite{sommese2020manycast2} wherein a prefix is advertised and reachable from multiple 
geographic regions.  We therefore utilize the current state-of-the-art
``AnyCatch'' technique and public service~\cite{anycatch}.
AnyCatch runs its own anycast instance, and issues ICMP probes
from different physical locations.  If the target is itself not
anycast, the ICMP responses should take the single shortest path back
to a single instance in the AnyCatch network.  However, if the target
is anycast, the responses will return to different nodes within
AnyCatch.  We ignore and do not geo-audit any prefixes that are
known to be anycast.
\end{itemize}

After identifying a likely responsive target within a prefix 
to geo-audit, we
instruct Atlas to issue ICMP measurements from 20 different vantage
points as depicted in Figure~\ref{fig:atlas}: 
three (3) vantages within the region of each of the five (5)  RIRs, and 
five (5) vantages within the country to which the prefix is registered.  Each
measurement is a one-off measurement, public, and tagged to
preserve the measurement artifact.  We then asynchronously fetch
results from Atlas and insert them into our database.

\begin{figure}[t!] 
 \centering
 \resizebox{0.7\columnwidth}{!}{\includegraphics{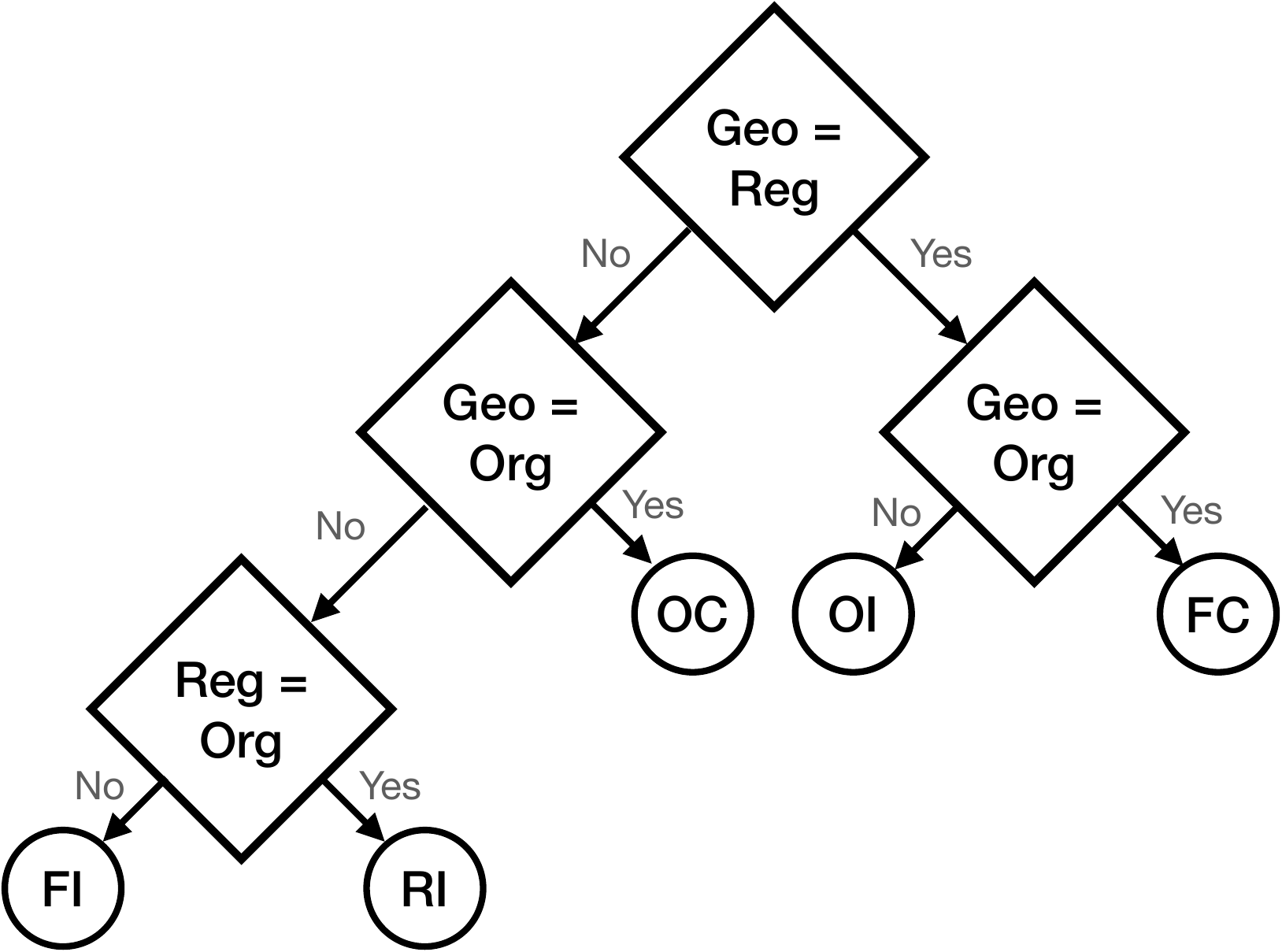}}
 \caption{Classification: Given a prefix allocated by
 $RIR_{Reg}$, with an organization in $RIR_{Org}$, we determine
 the RIR corresponding to the prefixes' inferred geolocation
 $RIR_{Geo}$ (\S\ref{sec:method:delay}). We then compare the three RIR's to 
 classify the prefix as either: 
(FC) Fully Geo-consistent; 
(OC) Organization Geo-consistent;
(OI) Organization Geo-inconsistent;
(RI) Registry Geo-inconsistent; or
(FI) Fully Geo-inconsistent.
 }
 \label{fig:classify}
\end{figure}

\subsection{Delay-based RIR Inference}
\label{sec:method:delay}

As detailed previously in \S\ref{sec:interregion}, the organization
to which a prefix is allocated may have a mailing address outside of
the RIR that manages the prefix's supernet.  For a prefix $P$ 
allocated to organization $Org$ by $RIR_{Reg}$, where the contact information
for $Org$ lists an address in country $CC$, we
define three important audit values: $RIR_{Reg}$ is the RIR in
which the prefix is registered; $RIR_{Org}$ is the RIR responsible
for $CC$, and $RIR_{Geo}$ is the RIR for $P$ based on
delay-based geolocation.

From the 20 different vantages, Atlas performs 3 ICMP echo requests to
the target.  Among the responses (up to 60 ICMP echo replies) we find
the vantage returning the minimum RTT to the target.  We then set
$RIR_{Geo}$ according to the RIR responsible for the country in which
this Atlas node is located.  Note that while latency-based geolocation has
been shown to be inaccurate for precision geolocation, conversely it
provides a much higher degree of accuracy at the
country-level~\cite{2011-huffaker-gt}.  In
this work, we use latency-based geolocation at an even coarser
granularity -- continents and RIR regions.  Further, we conservatively 
rely on
latencies to show that resources are \emph{not} in a particular RIR
region. 

\subsection{Taxonomy of Prefix Registration Geo-Consistency}

Given these three audit values, where each can be one of five
different RIRs, there are five different possible inferences;
table~\ref{tab:terminology} summarizes these possibilities and
provides an example for each.  
A ``Fully Geo-consistent'' result is one where the registered RIR, RIR
of the organization's country code, and the RIR of the nearest Atlas
probe all match.  A ``Organization Geo-consistent'' result is one where the
prefix's RIR is different than the RIR of the organization's mailing
address, but the inferred geolocation matches the organization's
country code RIR.  These first two results indicate that the prefix is
geo-consistent -- \ie that the audit passes.

The three remaining results show some form of unexpected
inconsistency.  For instance ``Organization Geo-inconsistent'' occurs when
our inferred RIR geolocation matches the RIR of the prefix, but not
the organization's responsible RIR.  Or, the inferred geolocation may
indicate an RIR different from both the registered RIR and registered
country code RIR, but the registrations belong to the same RIR, a
result we term ``Registry Geo-inconsistent.'' Finally, if all three
values are different, we term the result ``Fully Geo-inconsistent.'' 
Algorithm~\ref{alg:audit} in Appendix A provides the inference
pseudocode in additional detail.

%% file: results.tex
\ifx
\begin{table*}[]
{\small 
\resizebox{1.7\columnwidth}{!}{
\begin{tabular}{lrrrrr} 
Result              &      ARIN & RIPE & APNIC & LACNIC & AFRINIC \\
\midrule 
Fully Geo-consistent      & 94.7\% & 98.1 & 98.1\% & 97.0\% & 81.3\% \\
Organization Geo-consistent    & 1.2\% & 1.1\% & 0.5\% & 0.8\% & 7.6\% \\
\midrule 
Organization Geo-inconsistent  & 0.8\% & 0.2\% & 0.2\% & 0.0\% & 0.0\%\\
Registry Geo-inconsistent & 3.2\% & 0.4\% & 1.1\% & 2.1\% & 10.2\%\\
Fully Geo-inconsistent    & 0.1\% & 0.2\% & 0.1\% & 0.0\% & 0.9\% \\
\bottomrule
\end{tabular}}} 
\end{table*}
\fi

\section{Geo-Audit Results}
\label{sec:results}

In this preliminary work, we randomly select 10k prefixes from each of
the five RIRs for geo-audit.  Thus, we instruct Atlas to send 
3M ICMP echo requests (50k x 20 vantages x 3 packets) over a 10
day period from May 7-17, 2023.  For transparency and reproducibility,
all 50k of our measurement result sets are publicly available from Atlas
and easily found via the 
``\texttt{neo-ip-20230517}''
tag.  Each
measurement's meta-data encodes the RIR prefix, date, and other
relevant information.

\begin{figure}[t!] 
 \centering
 \resizebox{1.0\columnwidth}{!}{\includegraphics{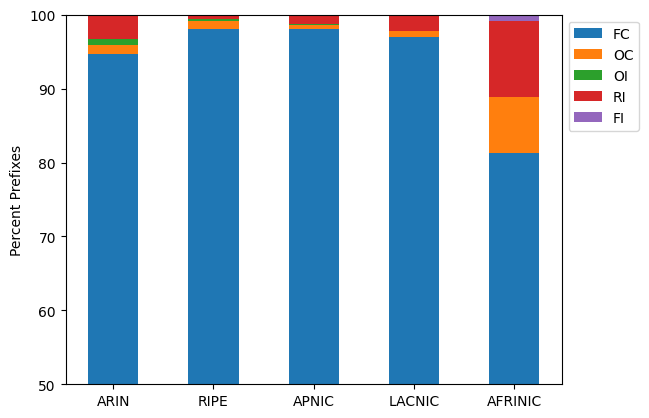}}
 \vspace{-2mm}
 \caption{Per-RIR geo-audit results: Over 96\% of 
classified prefixes are fully or organization consistent (FC and OC).
RIPE is the most consistent, while AFRINIC the least with 
more than 10\% registration inconsistent (RI).  Further, AFRINIC
leads in fully inconsistent (FI) prefixes.
}
 \label{fig:distrib}
\end{figure}

Among these 50k prefixes, we classify 44,936 according to the 
method of Figure~\ref{fig:classify} and taxonomy of 
Table~\ref{tab:terminology}. We ignore 18 prefixes that
belonged to a known anycast network.  For
the remaining 5k prefixes, our active probing was unable to elicit
enough responses (\eg due to firewall filters blocking ICMP
probes, or a non-responsive target in the hitlist) 
from the target to make a reliable inference.
Table~\ref{tab:aggresults} presents the aggregate geo-audit results,
while Figure~\ref{fig:distrib} shows the per-RIR classification
distribution.

\begin{table}[t] 
\begin{tabular}{lr} 
Result                    &    Prefixes \\
\midrule 
Fully Geo-consistent      &    42,270 (93.9\%) \\ 
Organization Geo-consistent    &    1,001 (2.2\%) \\
\midrule 
Organization Geo-inconsistent  &    120 (0.3\%) \\
Registry Geo-inconsistent &    1,527 (3.4\%) \\
Fully Geo-inconsistent    &    114 (0.3\%) \\
Anycast                   & 18 \\
\bottomrule
\end{tabular}
\caption{Aggregate Geo-audit Results}
\label{tab:aggresults}
\end{table}

\begin{table*}[t] 
\caption{Per-RIR geo-audit results.}
\label{tab:results}
{\small 
\resizebox{1.7\columnwidth}{!}{
\begin{tabular}{lrrrrr} 
                    &      \multicolumn{5}{c}{RIR} \\ 
Result              &      ARIN & RIPE & APNIC & LACNIC & AFRINIC \\
\midrule 
Fully Geo-consistent      & 8,614 (94.7\%) & 8,883  (98.1\%) & 8,893 
(98.1\%) & 8,641 (97.0\%) & 7,239 (81.3\%) \\
Organization Geo-consistent    & 107 (1.2\%) & 96 (1.1\%) & 49 (0.5\%) & 71
(0.8\%) & 678 (7.6\%) \\
\midrule 
Organization Geo-inconsistent  & 77 (0.8\%) & 22 (0.2\%) & 20 (0.2\%) & 0 (0.0\%)
& 1 (0.0\%)\\
Registry Geo-inconsistent & 290 (3.2\%) & 39 (0.4\%) & 99 (1.1\%) & 189
(2.1\%) & 910 (10.2\%)\\
Fully Geo-inconsistent    & 8 (0.1\%) & 18 (0.2\%) & 5 (0.1\%) & 4
(0.0\%) & 79 (0.9\%) \\
\bottomrule
\end{tabular}}} 
\end{table*}

We observe that, as a relative proportion of
the prefixes audited, the registry information is largely
consistent: overall 96.1\% are either fully 
or organization geo-consistent.  RIPE exhibits the most
consistency (over 99.2\%).  
APNIC and LACNIC are also highly consistent, while ARIN is
measurably lower, but still mostly consistent (95.9\%).
Further, \afrinic exhibits markedly less geo-consistency
with only 81.3\% of the prefixes we audit being fully consistent
and more than 10\% are inferred to reside outside of the
region.  \afrinic also leads in fully geo-inconsistent
prefixes (nearly 1\%).  

Delving into the inconsistencies, we find that the primary contributor
(approximately 75\%) to ARIN geo-consistencies are prefixes that our
audit shows to be in Mexico.  Conversely, over 50\% of the LACNIC
geo-inconsistencies are prefixes that are physically in the United
States.  The geolocation of geo-inconsistent \afrinic prefixes is
strongly dominated by Europe, but also with an appreciable number
of prefixes that geolocate to ARIN and APNIC countries..

\subsection{Case Study}


As an example of registration inconsistency, we present just
one instance discovered during our RIR geo-audit.  As part of our
measurements, our methodology selected a target address in the 
prefix \texttt{144.24.0.0/16}.  While IANA delegates responsibility
for the /8 prefix to which this belongs to ARIN, the /16 was
transferred to RIPE in 2004 as part of the Early Registration
Transfer Project (ERX).  The intent of ERX was to transfer
management of address space registration that pre-dated the creation of 
the other regional 
registries to ``the applicable RIR according to the
region in which the resource holders reside.''~\cite{arinerx}


The RIPE registration for this prefix lists the prefix description as
``Oracle Corp UK,'' with a mailing address in Great Britain, while the
responsible organization is ``Oracle Svendska AB,'' with a mailing
address in Sweden.  Our probe logic instructed Atlas to send ICMP echo
requests from 20 different vantage points on May 12, 2023.  Because
the registration lists a UK mailing address, the algorithm selected
five Atlas probes in Great Britain. However, among the five responses
to the Great Britain nodes the minimum RTT to the target was 129ms --
a latency that would be surprisingly high if the target were also in
Great Britain.

The logic further selected three probes in Kenya that produced a
minimum RTT of 258ms, indicating that the target is likely not in
Africa.  Similarly, the logic chose Brazil to test for presence in the
LACNIC region, and Korea for the APNIC region.  It further sent probes
from Germany to test whether it was anywhere in the RIPE region,
however the German probes returned a minimum RTT of 149ms, a value
even larger than found for the Great Britain probes.  Last, our logic
selected probes in Canada to test the ARIN region, which produced a
minimum RTT of 71ms.  Among the 20 probes the Canadian probe produced
the lowest RTT.  Upon further refinement, nodes in the US produced
even smaller RTTs and we found via additional probing a minimum RTT of
14ms using a probe in the state of Arizona.  Oracle runs a hosting
center in the Phoenix, AZ area, and we are confident that the prefix
is neither physically present in Great Britain, Sweden, or Europe. 
 
Per our audit rules, the algorithm labeled this prefix as ``registry
geo-inconsistent:'' while the mailing addresses of the registrant are
in Europe and therefore are under the purview of RIPE, the
geo-location is elsewhere.  It is unclear why this organization is
using address space managed by RIPE and registered to the European
branch of the company to number resources that are physically in the
United States.

\subsection{Limitations}
\label{sec:results:limits}

There are several potential limitations with our methodology.  First,
we may be unable to find an ICMP-responsive target within a particular
prefix, which prevents an active measurement location audit.  Second, 
Atlas may have no probe vantages within the country where a prefix
is registered.  If we are unable to obtain in-country results, we 
conservatively make no audit determination.  

Third, the reported
location of the Atlas nodes may be incorrect.  While Atlas introduces
a small amount (1km) of error into the location of probes, this error
will not materially affect our inferences.  However, the probe may be
in a physically incorrect location.  We attempt to mitigate an
erroneously located probe by requesting five different in-country
probes; since we take the minimum RTT, a probe incorrectly listed
as in-country will simply be ignored for the purposes of computing
the region.  Nonetheless, to ensure the reliability of our inferences,
we performed a second ``refinement'' round of probing wherein we 
re-measured with different in-country probes those prefixes where
we found geo inconsistencies.  This second-round of probing did not
alter our findings, hence we believe our geo-inferences to be largely
correct.

Finally, we focus exclusively on IPv4 registrations and leave analysis
of IPv6 to future work. 

%% file: concl.tex
\section{Conclusions}
\label{sec:conclusions}

This work takes the first steps towards understanding the geographic
allocation and use of IPv4 addresses.  In future work, we plan to
geo-audit all $\sim$8M prefix registrations and extend the survey to
audit IPv6 allocations.  The long history of RIRs and inter-RIR
transfers suggests that the older RIRs may have an accumulation of
inaccuracies, while \afrinic's legacy of colonialism and connectivity
via Europe may in part explain some of the inconsistencies we see.
This initial work seeks only to measure inconsistencies -- going
forward, we wish to engage with the RIRs to validate and better understand the
causes for the inconsistencies we uncover

We hope these results can meaningfully contribute to the important
policy and technical discussions surrounding IP address allocation, as
well as highlight areas where RIRs may wish to work with their
constituent membership to improve registration accuracy.

%% file: alg.tex
\section{Algorithm}

We provide pseudo-code for our geo-audit inference algorithm here for
additional detail.  The algorithm assumes a database operation,
db\_lookup, (as described in our methodology) that performs a
longest-prefix match on a prefix $P$ to return the RIR to which it is
registered.  It further assumes (again, as described in our
methodology) a lookup function, rir\_lookup,  between an ISO country
code and the responsible RIR.  Finally, we abstract the RIPE Atlas
probing logic and assume a function that returns the set of ICMP
RTTs as produced by our active measurements in the atlas\_results
function.

The algorithm returns one of six possible results: anycast ($AC$), 
Fully Geo-consistent ($FC$), Organization Geo-consistent ($OC$), 
Organization Geo-inconsistent ($OI$), Registry Geo-inconsistent ($RI$)
or Fully Geo-inconsistent ($FI$).  Section~\ref{sec:method} details
each step of our audit methodology, while Section~\ref{sec:results}
provides results from a 50k prefix RIR geo-audit.

\begin{algorithm}[]
\caption{GeoAudit($P$)}
\label{alg:audit}
\begin{algorithmic}
\If {$P\in anycatch$}
  \State return($AC$)
\EndIf
\State $RIR_{Reg}\gets$ db\_lookup($P$)
\State $CC\gets$ db\_lookup($P$)
\State $RIR_{Org}\gets$ rir\_lookup($CC$)
\State $RTT[]\gets$ atlas\_results($target \in P$)
\State $Probe_{min} = \underset{k}{\mathrm{arg\_min}}$ $RTT[k]$ 
       $\forall k\in probes$
\State $RIR_{Geo}\gets$ rir\_lookup($Probe_{Min}$)
\If {$RIR_{Geo} = RIR_{Reg}$}
  \If {$RIR_{Geo} = RIR_{Org}$}
    \State return($FC$)
  \Else
    \State return($OI$)
  \EndIf
\Else
  \If {$RIR_{Geo} = RIR_{Org}$}
    \State return($OC$)
  \ElsIf {$RIR_{Reg} = RIR_{Org}$}
    \State return($RI$)
  \Else
    \State return($FI$)
  \EndIf
\EndIf
\end{algorithmic}
\end{algorithm}

